\documentclass[10pt]{article}
\usepackage{amsfonts}
\usepackage{epsfig}
\usepackage{amssymb,amsmath,bm}
\usepackage{subfigure}
\usepackage{amsthm}
\usepackage{mathrsfs}
\usepackage[section]{algorithm}
\usepackage{fancyhdr}

\begin{document}

\begin{center}

{\Large \textbf{Robust and Adaptive Algorithms for Online Portfolio Selection}}

Theodoros Tsagaris\footnote{GSA Capital London.
The views presented here reflect solely the author's opinion. Work completed
whilst a PhD student at Imperial College London.}, 
Ajay Jasra\footnote{Imperial College London} \& Niall Adams\footnote{Imperial College London}

\end{center}


\begin{abstract}
We present an online approach to portfolio selection.
The motivation is within the context of algorithmic trading,
which demands fast and recursive updates of portfolio allocations, as new data arrives.
In particular, we look at two online algorithms: Robust-Exponentially Weighted
Least Squares (R-EWRLS) and a regularized Online minimum Variance algorithm (O-VAR). Our methods use simple ideas from signal processing and statistics, which
are sometimes overlooked in the empirical financial literature.
The two approaches are
evaluated against benchmark allocation techniques using 4 real datasets. Our
methods outperform the benchmark allocation techniques in these datasets, in terms of both computational demand and financial performance.
\end{abstract}



\section{Introduction}

 In portfolio allocation problems, investors aim to optimize the return of the invested capital based on some cost function by allocating a fraction of the capital in a number of different assets. In the long established \textit{mean-variance} theory (see Markowitz (1952)) for asset allocation, the fraction of the capital invested in each asset is known as the portfolio weight, and all weights together form a linear combination (portfolio) that is optimal when the expected return of the portfolio is maximized for a fixed level of variance of the portfolio. 
The approach argues that maximization of expected returns
does not guarantee that the portfolio will have the smallest
variance. Hence, a trade-off between the expected return
and the variance of the portfolio provides a more effective diversification
of investors funds. Investors are considered risk averse and would prefer the portfolio with the smallest risk when expected returns are equal. Moreover, a portfolio with smaller variance is a desirable attribute, as investors could leverage  by increasing the capital allocation, so that the portfolio would achieve higher return on capital.
Although the mean-variance analysis theory, initially, generated little interest, it is now a mainstream theory whose principles are constantly visited and re-invented. 
We also wish to clarify that the meaning of the terms \textit{assets} and \textit{instruments} are used in this text interchangeably and they are deemed as available investment vehicles.

However, in mean-variance optimization it is well known that the portfolio weights
can be highly unstable. This is due to the difficulty of estimating expected
returns; see Merton (1980). As a result, there has been a substantial amount
of recent interest in improving estimation procedures, including: Baltutis (2009); DeMiguel \& Nogales (2009); DeMiguel et al.~(2009a,b); Fabozzi et al.~(2007); Fabozzi et al.~(2009); Jagannathan \& Ma (2003); Ledoit \& Wolf (2003, 2004).
This work ranges from imposing constraints on the optimization function,
to robust portfolio estimation procedures. Whilst these publications are vital in the
understanding of portfolio allocation, they are mainly concerned with \textit{batch
procedures} --that requires historical observations-- as opposed to \textit{online} techniques which are equipped with  recursive estimation mechanisms. Batch procedures are  not necessarily designed to be computationally
efficient and address the streaming nature of financial data, nor to handle the high dimensionality of the available assets for allocation.

We approach the asset allocation problem from the algorithmic trading perspective, that is when investment decisions regarding allocations are taken automatically through investment allocation algorithms, as soon as data arrive.  
\textit{Algorithmic trading}, otherwise known as automated or systematic trading, refers to
the use of algorithms to conduct trading without any human intervention.
As an example, in 2006, one-fifth of global equity trading was administered through
algorithmic techniques (Keehner (2007)). Such transactions are executed within a few milliseconds and any latency can make a difference between a profitable or loss making trade. 
In this context batch algorithms are unsuitable and we must consider online procedures.  One such consideration is implementation of algorithms relating to portfolio optimization.

In this study we use ideas from mean-variance theory to automate the  process regarding portfolio optimisation.
In particular, we make use of the algebraic link of the classic mean-variance theory to Ordinary Least Squares (OLS) to allocate capital among various assets.
We construct these algorithms bearing in mind certain considerations with regard to efficiency of trading and characteristics of financial data.
These algorithms may account for one or more of the following attributes:

\begin{itemize}
\item \textit{Adaptive}: They have the ability to adapt to non-stationary market environments.  By dynamically
incorporating new information into the portfolio weights, one is likely to improve
the financial performance of the resulting algorithms.
\item \textit{Robust}:  They are able to counter the adverse effect of  outliers in estimation.
\item \textit{Regularized}: They  have mechanisms to reduce the high level of noise exhibited in financial data, either though direct regularization or dimensionality reduction techniques. 
\item \textit{Efficient}: 
They are sequential, one-pass methods to suit to the nature of the problem, that is to process information fast in order to exploit investment opportunities as they occur.
\end{itemize} 

The above considerations, together with ideas of asset allocation using regression, enables us to devise  suitable techniques for algorithmic trading. We will use these techniques on real datasets and compare them against established and well documented asset allocation methods.

\subsection{Contribution and Structure}

Online or multi-period portfolio optimization has been
investigated in the literature, a non-exhaustive list includes: Agarwal et al.~(2006); Chapados (2007); Frauendorfer \& Seide (2000); Helmbold et al.~(1998);
Kuhn et al.~(2009); Li \& Ng (2000); Smith (1967). Montana et al.~(2008, 2009) investigated online algorithms for statistical arbitrage trading strategies.
Some of the (computationally fast) portfolio optimization techniques
originate in computer science/machine learning and they are algorithmically
distinct from the standard mean-variance type procedures that are often found
in the empirical finance literature (as exemplified in the list above).
As such, one of the main objectives of this article is to bridge efficient algorithmic techniques found in various disciplines  with long established
portfolio selection literature in finance. The online algorithms are developed here for three reasons.
\begin{itemize}
\item{For their importance from an applied perspective.}
\item{To cross fertilize financial ideas,  with ideas from signal processing,
statistics, computer science and lead to more efficient techniques.}
\item{To illustrate the potential improvements in financial performance.}
\end{itemize}

There are a substantial number of ideas in the listed financial literature which can
improve the current allocation techniques. However, they appear to be
seldom used in empirical finance and our objective is to provide a simple exposure to these ideas.
For example, the constraints typically used in mean-variance problems (e.g.~DeMiguel
et al.~(2009)) correspond to standard Tikhonov regularization and are well-understood in signal processing as helping to guard against instability induced by ill-conditioned matrices. Ill-conditioned matrices are often encountered in mean-variance theory because of the multi-collinearity of asset log-returns, which may lead to rank deficient problems (see Hansen (1996)). 
Moreover, as mentioned earlier, adaptive algorithms have the ability to adapt their estimates to the underlying data and they are naturally more suitable for non-stationary environments, such as those in finance.
In the sequel, we construct two algorithms which are related to batch mean-variance and minimum variance methodology. These two methods use simple ideas from signal
processing and statistics to construct fast and robust approaches to portfolio
selection.

This paper is structured as follows. In Section \ref{sec:Markowitz} the mean-variance
theory and our online framework is introduced. In Section \ref{sec:R-EWRLS} the computation for our two methods is developed.
In Section \ref{sec:application}
our methods are applied to 4 real datasets and, finally, in Section \ref{sec:conclusion} we conclude the paper discussing possible avenues of future work.


\subsection{Notation and Set-Up}
In this paper, the following notations are adopted.
All vectors are column vectors, and we denote the transpose by the prime symbol i.e. $(\beta')'=\beta$. 
The column vector of $d$ ones is written $\mathbb{I}_d$ and the $d\times
d$ identity matrix is written $\mathbb{I}_{d\times d}$.
Given a collection of $d-$vectors $x_j,\dots,x_T$,
$1\leq j<T$ say, the $(T-j+1)\times d$ matrix composed of the concatenation of these vectors is written $X_{j:T}$.
Denote the H\"{o}lder $p-$norm by
$\|\beta\|_p=(\sum_{i=1}^p |\beta_i|^p)^{1/p}$. The trace of a matrix $A$
is written $\textrm{tr}(A)$.

\section{Portfolio Selection}\label{sec:Markowitz}
In the following section we introduce the problem and describe our framework.
The log-returns of $d$ financial instruments are observed at times
$1,2,\dots,T$: $x_1,\dots,x_T$, $x_{n}=(x_{1,n},\dots,x_{d,n})'$ for $n\in\{1,\dots,T\}$. An investor seeks to construct a portfolio by
 optimally (in some sense) allocating funds to a collection of  $d$ instruments. 


\subsection{Batch Portfolio Selection}\label{sec:meanVarProblem}

Most portfolio selection problems are stated in a static or batch manner.
For completeness we describe the mean-variance theory (Markowitz
 (1952)). Denote the mean
and covariance matrices of the log-returns as $\mu$ and $\Sigma$ respectively.
Then the objective is to solve the problem
$$
\max_{\beta} \bigg\{\beta'\mu - \frac{1}{2}\beta'\Sigma \beta\bigg\}
\quad
\textrm{s.t.}~\beta'\mathbb{I}_d = 1,
$$
where $\beta$ is the $d$-vector of portfolio weights.
This optimization problem is straight-forwardly solved via Lagrange multipliers.
In practical situations,  the estimated mean and covariance is substituted into the optimization
problem, leading to a data-dependent solution.
Intrinsically, many of the portfolio optimization problems that are considered
in the literature
may be written as
$$
\max_{\beta} \bigg\{f(X_{1:T};\beta) + \eta[\beta'\mathbb{I}_d - 1]\bigg\}
\quad
\textrm{or}
\quad
\min_{\beta} \bigg\{f(X_{1:T};\beta) + \eta[\beta'\mathbb{I}_d - 1]\bigg\}
$$
for some function $f$, Lagrange multiplier $\eta$ and matrix of log-returns $X_{1:T}$. For example, one
of the problems in DeMiguel et al.~(2009b):
$$
\min_{\beta} \bigg\{\beta'\hat{\Sigma}\beta + \delta\|\beta\|_1 + \eta[\beta'\mathbb{I}_d - 1]\bigg\}
$$
corresponds to a minimum variance portfolio with $\mathbb{L}_1-$constraints, where that ``hat'' notation refers to an estimated quantity. 
We note that this approach involves constructing a covariance matrix and subsequently computing its inverse to arrive to a solution. As mentioned earlier, $d$ can be very large and this often leads to computational delays. These computational delays can be detrimental in algorithmic trading, where tick data are streaming and decisions about allocation need to be taken instantly based on the latest information. 

Another known portfolio allocation technique, which is used throughout this article, is the \textit{naive} strategy which assigns equal constant portfolio weights to all instruments in the portfolio (i.e. $ \mathbb{I}_d1/d$). This simple allocation technique is of practical importance, as it has been shown in an empirical study by  DeMiguel et al.~(2009b) to outperform many more complicated allocation techniques.

\subsection{Online Portfolio Selection}\label{sec:online_optimization}
The simple extension that is studied in this paper, is to consider:
\begin{equation}
\min_{\beta_n} \bigg\{f_n(X_{\alpha_n:n};\beta_n) + \eta_n[\beta_n'\mathbb{I}_d - 1]\bigg\} \quad\quad \alpha_n = 1\vee(n-W+1)\quad
n\in\{1,\dots,T\}\label{eq:objective}
\end{equation}
where $\eta_n$ is a Lagrange multiplier and $W$ is a fixed window of data. That is, the parameters are now estimated over a sliding window $W$, rather using all available data. Note that when $W=1$, then $\alpha_n=n$ i.e. $X_{n:n}$ which is the vector $x_n$. This is chosen to ensure that our algorithms
are of approximately fixed computational complexity per time-step (see Section
\ref{sec:initialization} for discussion on window length selection).
Note, the larger the sliding window, the more data are used for estimation. Conversely, the smaller the sliding window, the more weight is given to more recent data. 
\eqref{eq:objective} includes
some interesting special cases such as:
\begin{equation}
f_n(X_{\alpha_n:n};\beta_n) = \|\mathbb{I}_W - X_{\alpha_n:n}\beta_n\|_2^2 + \delta_n \|\beta_n\|_2^2 \label{eq:objective_recursive}
\end{equation}
which could be considered a sequential ridge-regression, for $\delta$ being the regularization parameter. This latter formulation
is equivalent to a mean-variance problem (see Section \ref{sec:online_optimization}) with $\mathbb{L}_2-$constraints;
see Britten-Jones (1999) for details. Note also, that the function 
in Helmbold et al.~(1998) ($F$ in their notation) also falls into the framework above.

The reason for giving \eqref{eq:objective_recursive} is to provide a link between mean-variance theory and recursive estimation algorithms. As such, we are able to devise recursive asset allocation algorithms, through the use of recursive least squares, for dealing with streaming data and take advantage of the number of regularisation methods developed for  regression  to deal with the inherent instability of  the portfolio solution to estimation error. 

\subsubsection{Objective Functions} 

The first case we propose is 
\begin{equation}
f_n(X_{\alpha_n:n};\beta_n) = \sum_{i=1}^n \lambda^{n-i}\rho(r_i(\beta_n))
\label{eq:sestimation}
\end{equation}
with $W$ equal to the size of all available observations, $\rho:\mathbb{R}\rightarrow\mathbb{R}^+$ differentiable and
$
r_i(\beta_n) = \sum_{j=1}^{d}\big[(1-x_{j,i}\beta_{j,n})/\sigma_{j,i}\big].
$
The parameter $\sigma$ is a scale parameter estimate that is used to standardize the residual error $(1-x_{j,i}\beta_{j,n})$; we use a robust scale parameter defined later in Section \ref{sec:robRecScaleEstim}.
The parameter $\lambda$ is a forgetting factor; this is a well-known tool in adaptive filtering e.g.~Haykin (1996). The choice of 1 in $r_i(\beta_n)$ follows the work in Britten-Jones (1999).
A heuristic explanation is as follows: setting the response variable equal to a positive constant implies that our portfolio is minimised against an ideal portfolio  that has  positive returns for each timestep and is risk-less (a vector a constant has zero variance).

The objective function \eqref{eq:sestimation} corresponds to a sequential
form of M-estimation (see e.g.~Deng (2008) for related ideas). 
\eqref{eq:sestimation} follows
the recent trend in portfolio optimization to use robust statistical procedures
to estimate parameters of interest; see e.g.~DeMiguel \& Nogales (2009).
For reasons that will become apparent, the approach associated to \eqref{eq:sestimation} is termed robust-exponentially weighted recursive least squares (R-EWRLS).

The second case is:
\begin{equation}
f_n(X_{\alpha_n:n};\beta_n) = \frac{1}{2}\bigg[\beta_n' (X_{\alpha_n:n}'F_{\lambda_n}X_{\alpha_n:n})\beta_n
+ \delta_n \|\beta_n\|_2^2\bigg]
\label{eq:minvarl2}
\end{equation}
where $F_{\lambda}=\textrm{diag}(\lambda^{W},\lambda^{W-1},\dots,1)$.
The task of estimating $\lambda$ and $\delta$ parameters is discussed later in Section \ref{sec:adaptivedelta}. This corresponds to an online minimum-variance-type algorithm with $\mathbb{L}_2-$constraints
(termed online minimum-variance (O-VAR) throughout).
The matrix $F_{\lambda}$ introduces a forgetting-factor into the optimization
scheme.
The use of the estimated second moment, instead of the covariance is for computational
reasons; we did not find a substantial discrepancy (in terms of financial
performance) when compared to using the covariance matrix. Note that a more
standard recursive estimate could be obtained using the function
$$
f_n(X_{1:n};\beta_n) = \frac{1}{2}\bigg[\beta_n' (x_{n}'x_{n})\beta_n
+ \delta_n \|\beta_n-\beta_{n-1}\|_2^2\bigg]
$$
but is not considered here, due to the relationship of \eqref{eq:minvarl2} to the standard minimum-variance
approach.

The batch version
of \eqref{eq:minvarl2} is studied in DeMiguel et al.~(2009b). The $\mathbb{L}_2-$constraints correspond to an $\mathbb{L}_2$ distance with the naive allocation
strategy. The naive approach to allocation surpasses estimation of the sample mean and one would
expect relatively stable portfolio weights.

Note that, for both procedures there are unknown parameters $\lambda$, $\delta$
and $\sigma$. The next section discusses how these parameters may be set, in addition to recursive formulation of the proposed optimizations.


\section{Updating Schemes}\label{sec:R-EWRLS}

In this section, we introduce our recursive updating approaches. This section is core to the development of the adaptive allocation algorithms as it formulates efficient regression techniques appropriate to the nature of algorithmic trading.

\subsection{R-EWRLS}\label{sec:seqsestimation}

Let us introduce some notations: 
$$
\widetilde{x}_n = (x_{1,n}/\sigma_{1,n},\dots,x_{d,n}/\sigma_{d,n})'\quad\quad \bar{\sigma}_n =  \sum_{j=1}^d\bigg[\frac{1}{\sigma_{j,n}}\bigg]\quad\quad
q(x)  =  \frac{1}{x}\frac{d\rho}{d x}(x).
$$
Then, ignoring the Lagrange multiplier (the result can be renormalized),
we are to minimize \eqref{eq:sestimation}. Differentiating, it follows that
the optimal $\beta_n$ solves
$$
\sum_{i=1}^n\lambda^{n-i}q(r_i(\beta_n))\bar{\sigma}_i\widetilde{x}_i = 
\sum_{i=1}^n\lambda^{n-i}q(r_i(\beta_n))\bigg[\sum_{j=1}^d\frac{x_{j,i}}{\sigma_{j,i}}\beta_{j,n}\bigg]\widetilde{x}_i.
$$
Since this equation is often non-linear, we use the approximation $r_n(\beta_n)
=r_n(\beta_{n-1})$, with $\beta_{n-1}$ given (i.e.~by the previous step, or
by initialization).
Now, let $z_n$ denote the L.H.S.~and $\Phi_n = \sum_{i=1}^n \lambda^{n-i} q(r_i(\beta_n))\widetilde{x}_i\widetilde{x}_i'$, then we are to solve
$$
z_n = \Phi_n \beta_n.
$$
As $\Phi_n = \lambda\Phi_{n-1} + q(r_n(\beta))\widetilde{x}_n\widetilde{x}_n'$,
and writing $P_n=\Phi_n^{-1}$, it follows via the Sherman-Morrison (e.g.~Haykin (1996)) formula
$$
P_n = \lambda^{-1}P_{n-1} - \lambda^{-1}\kappa_n\widetilde{x}_n' P_{n-1}
$$
with
$$
\kappa_n = \frac{ q(r_n(\beta_{n-1}))\lambda^{-1}P_{n-1}\widetilde{x}_n}
{1+q(r_n(\beta_{n-1}))\widetilde{x}_n' P_{n-1}\widetilde{x}_n}.
$$
Using $z_n = \lambda z_{n-1} + q(r_n(\beta_n))\bar{\sigma}_n \widetilde{x}_n$ we thus have the recursion
$$
\beta_n = \beta_{n-1} + q(r_n(\beta_{n-1}))\bar{\sigma}_n P_n \widetilde{x}_n
- \kappa_n \widetilde{x}_n'\beta_{n-1}.
$$

We have presented a recursive least squares procedure whose algebraic equivalence with the Kalman filter is well-known and understood (see Chapter $12$ of Sayed (2003)).
It should be remarked that related ideas have appeared in Cipra \& Romera
(1991) and our approach is similar to  robust filters 
(Martin (1979); Masreliez (1975); Schick \& Mitter (1994)).


\subsubsection{Robust Recursive Scale Estimate}\label{sec:robRecScaleEstim}

The calculation of the scale parameter is now detailed. Our approach uses robust statistics. 
First, we note that the Median Absolute Deviation (MAD) (e.g.~Huber (2004)) estimate
of scale is given by
$$
MAD_{V}(X_{n-V+1:n})= \textrm{med}_{j}(|x_{i,j}-\textrm{med}_{l}(x_{i,l})|)\quad
\quad j,l\in \{1\vee n-V+1,\dots,n\}, i\in\{1,\dots,d\}
$$
where $V$ is a chosen data window and $\textrm{med}(\cdot)$ is the median function. 
Recent research has pointed to efficient techniques to compute the median with $O(V)$ average complexity using recursive binning schemes (see Tibshirani (2008)).

Second, an exponentially recursive median absolute deviation (EWMAD) estimator
is considered
\begin{equation*}
\widehat{\sigma}_{i,n}^{(\textrm{med})}=\nu\widehat{\sigma}_{i,n-1} +
c (1-\nu) \textrm{med}_{j}(|x_{i,j}-\widehat{\mu}_{i,n}^{(\textrm{med})}|) \quad j\in\{1\vee n-V+1,\dots,n\}
\end{equation*}
and where $\nu$ is another forgetting factor and
$\widehat{\mu}_{i,n}$ is an EWMED (Exponentially Weighted Recursive Median),
given by
\begin{equation*}
\widehat{\mu}_{i,n}^{(\textrm{med})}=\nu\widehat{\mu}_{i,n-1}^{(\textrm{med})}  + (1-\nu) \textrm{med}_{j}(x_{i,j})
 \quad j\in\{1\vee n-V+1,\dots,n\}
\end{equation*}
where $c=1/\Phi^{-1}(3/4)\simeq 1/0.6745$ 
is a correction factor to make MAD consistent with the normal distribution (e.g.~Huber (2004)). The EWMED is similar to the well documented EWMA (e.g.~Hamilton
(1994))
with the only difference that the EWMED estimator replaces the latest information $x_n$ by its median estimate over the sliding window. On the basis of much preliminary investigation on specific datasets, we have arbitrarily set $V=20$ and $\nu= 0.99$ for all of the applications. Due to the robust nature of
the above estimation, this method is termed robust-exponentially weighted
recursive least squares.

\subsubsection{Dealing with Noisy Data}\label{sec:low_rank_approx}

As discussed earlier, financial data are inherently noisy and exhibit high degree of dependence. The noise hampers the ability to accurately forecast and the dependence structure of assets accentuates the problem, as pointed out in the introduction; this is via the instability of portfolio weights caused by potential rank deficiency. To alleviate for these problems we adopt a low rank matrix approximation of  $X_{\alpha_n:n}$,
$W<\infty$ in order to eliminate those components of data that contain most of the noise. This approach aims to optimally approximate, with respect to some norm, a matrix of lower rank while retaining the same same dimension. It is well known that the best low rank approximation can be found by Singular Value Decomposition (SVD) under the Frobenius norm (see e.g.~Stewart (1993)).
The approach is as follows.

Let $1\leq \widetilde{r}< n \wedge d$ be given and denote
the singular value decomposition (SVD) of the returns matrix $X_{\alpha_n:n}=U_n\Gamma_n
V_n'$. Consider the truncated SVD (see Hansen (1987))
$$
\Gamma_{\widetilde{r}n}=\left(\begin{array}{cc} 
K_n & 0 \\ 
0 & 0\\ 
\end{array}\right)
\quad
K_n=\left(\begin{array}{ccc} \gamma_{1,n} & 0 & \cdots \\ 0 & \ddots & \cdots\\ \vdots & \vdots & \gamma_{\widetilde{r},n}\\ 
\end{array}\right)
$$
then set $\hat{X}_{\alpha_n:n}=U_n \Gamma_{\widetilde{r}n} V_n'$. We replace $x_n$ in the recursions in Section \ref{sec:seqsestimation} with the final row of $\hat{X}_{\alpha_n:n}$. The value of $\widetilde{r}$ is set during
training. Note that the SVD of $X_{\alpha_n:n}$ can be updated incrementally using the methods in (Bunch \& Nielsen (1978)).

\subsection{Online Minimum-Variance}

The minimum-variance scheme is somewhat less involved. Suppose $\lambda$,
$\delta_n$ and $W$ is given. It is straight-forward to show that, at time
$n$, the solution of the optimization problem \eqref{eq:objective}, with $f_n$ as in \eqref{eq:minvarl2} is
\begin{equation}
\beta_{n} = \frac{\bigg((X_{\alpha_n:n}'F_{\lambda}X_{\alpha_n:n}) + \delta_n\mathbb{I}_{d\times d}\bigg)^{-1}\mathbb{I}_d}
{\mathbb{I}_d'\bigg((X_{\alpha_n:n}'F_{\lambda}X_{\alpha_n:n}) + \delta_n\mathbb{I}_{d\times d}\bigg)^{-1}\mathbb{I}_d}
\label{eq:betaovar}.
\end{equation}
The main objective here is to calculate this quantity quickly. Suppose we
are given the eigen-decomposition of $X_{\alpha_n:n}'F_{\lambda}X_{\alpha_n:n}$,
i.e.~$X_{\alpha_n:n}'F_{\lambda}X_{\alpha_n:n} = Q_n \Pi_n Q_n'$, then the inverse
in \eqref{eq:betaovar} is equal to
$$
\bigg((X_{\alpha_n:n}'F_{\lambda}X_{\alpha_n:n}) + \delta_n\mathbb{I}_{d\times d}\bigg)^{-1} = Q_n(\Pi_n + \delta_n\mathbb{I}_{d\times d})^{-1}Q_n'
$$
that is, one need only calculate the inverse of a diagonal matrix. The recursive
calculation of the eigen-decomposition can be achieved by the methods
of Yu (1991) in $O(2d^2)$; i.e.~this operation is $O(d^2)$ instead of the
standard $O(d^{2+\theta})$ ($\theta>0$) for matrix inversion. More specifically,
the method of Yu (1991) is to re-calculate the new eigen-decomposition of
$R'$, from 
$R$ to $R'$ of  the form
$$
R' = R + \xi_1\xi_1' - \xi_2\xi_2'
$$
with $\xi_1,\xi_2$ vectors of the appropriate dimension.
In our case we have that
$$
\lambda X_{\alpha_{n+1}:n+1}'F_{\lambda}X_{\alpha_{n+1}:n+1}
= X_{\alpha_n:n}'F_{\lambda_n}X_{\alpha_n:n} - \lambda^W x_{\alpha_n}
x_{\alpha_n}' + \lambda^{-1} x_{n+1}x_{n+1}'
$$
so the same ideas may be applied. Note that the incremental SVD mentioned above could also be used.

\subsubsection{Adaptive Calculation of $\delta_n$ and $\lambda$}\label{sec:adaptivedelta}

There are still 2 free parameters to be set; $\delta_n$ and $\lambda$.

First, consider $\delta_n$. Lacking an analytical solution, we investigate $\delta$ numerically based on an initial training data period. To investigate the effect of $\delta$ perturbations to  portfolio returns, we choose  a short initial
training sequence of data to calculate
$\textrm{trace}(X_{\alpha_n:n}'F_{\lambda}X_{\alpha_n:n})$ for a given $\lambda$.
Then,  we select a collection of $G$ equally spaced points between 
$\textrm{tr}(X_{\alpha_n:n}'F_{\lambda}X_{\alpha_n:n})/d$ and
$\textrm{tr}(X_{\alpha_n:n}'F_{\lambda}X_{\alpha_n:n})$. The algorithm
is initialized at any of those points. At re-balancing times (the times when the allocation is altered) 
we compute the portfolio returns over the training period
for each of the $G$ points and select the one that generates the  largest portfolio return.
The range of the grid is based upon the recommendations in Ledoit \& Wolf
(2004). We found our results to be extremely robust to the initial value
of $\delta$.

Second, consider $\lambda$. In this scenario, we only recalculate $\lambda$ at
re-balancing times, which incurs the cost of re-computing the eigen-decomposition
of $X_{\alpha_n:n}'F_{\lambda}X_{\alpha_n:n}$. We follow a similar procedure to that
in adaptive filtering. An attractive criterion for portfolio selection,
is to minimize
$$
\|\mathbb{I}_{W} - X_{\alpha_n:n}\beta_n\|_2^2
$$
see Britten-Jones (1999). As a result, at the $m^{\textrm{th}}-$re-balancing time,
the following stochastic approximation type update is used:
$$
\lambda_{m} = \lambda_{m-1} + \frac{1}{ml}\sum_{j=(m-1)l + 1}^{ml}
\textrm{sgn}\bigg\{\frac{\partial}{\partial \lambda}\bigg[\|\mathbb{I}_W-F_{\lambda}X_{\alpha_j:j}\beta_j\|_2^2
\bigg]
\bigg\}.
$$
See e.g.~Chapter 14, Haykin (1996) for similar self-tuning approaches for recursive filtering. Note that 
if $\lambda_m \notin (0,1)$, then we set $\lambda_m = \lambda_{m-1}$.

\subsection{Discussion}

The two methods described here have some  complementary aspects. Firstly, from
the perspective of dealing with noisy data, the methods use separate, but
well known procedures. R-EWRLS uses the truncated SVD, whilst the
O-VAR uses a form of Tikhonov regularization via 
$\mathbb{L}_2-$constraint.
Secondly, the R-EWRLS method accounts for outliers by down-weighting them through a by-product weighting quantity ($q$, see Section \ref{sec:seqsestimation}) of the robust cost function. On the other hand, O-VAR does not have an embedded mechanism to account for outliers as they occur.

Thirdly, the O-VAR is adaptive to non-stationary environments and accounts for variability in the underlying environment through the self-tuning forgetting factor $\lambda$. However, the rank $\delta$ needs to be set during training. In R-EWRLS case, $\lambda$ needs to be calibrated in advance and such calibration needs to take place every time a  shift occurred in the underlying environment. Also, rank $\widetilde{r}$ of the low rank approximation (Section \ref{sec:low_rank_approx}) needs to be set in advance.  

It is likely that one procedure is likely to be preferred given the scenario.
For example, when the data are subject to a change in the economic cycle,
one would expect the O-VAR to perform significantly better, however, O-VAR it does not take into consideration expected returns.
In that respect, O-VAR may be more suitable for assets that are expected to grow in the future. For instance, it may be suitable for fund of funds whose underlying investments have positive expectation and desire to allocate robustly. Alternatively, it could be suitable for an algorithmic trading system that allocates between allocation strategies in an adaptive and efficient way.  
Finally, the R-EWRLS is linked to mean-variance theory and should be suitable for any asset class and as a standalone allocation strategy. Note that O-VAR is similar to a more efficient version of the function in DeMiguel et al.~(2009b).

\section{Application}\label{sec:application}

The techniques described in Section \ref{sec:R-EWRLS} 
are applied to 4 datasets. Financial performance
is compared to standard methods. Note that a zero-rate
risk free interest rate is assumed throughout.

\subsection{Data Description}\label{sec:data}
 
\begin{figure}
\centering
\subfigure[FX]
{\includegraphics[width=0.48\textwidth,height=4.2cm]{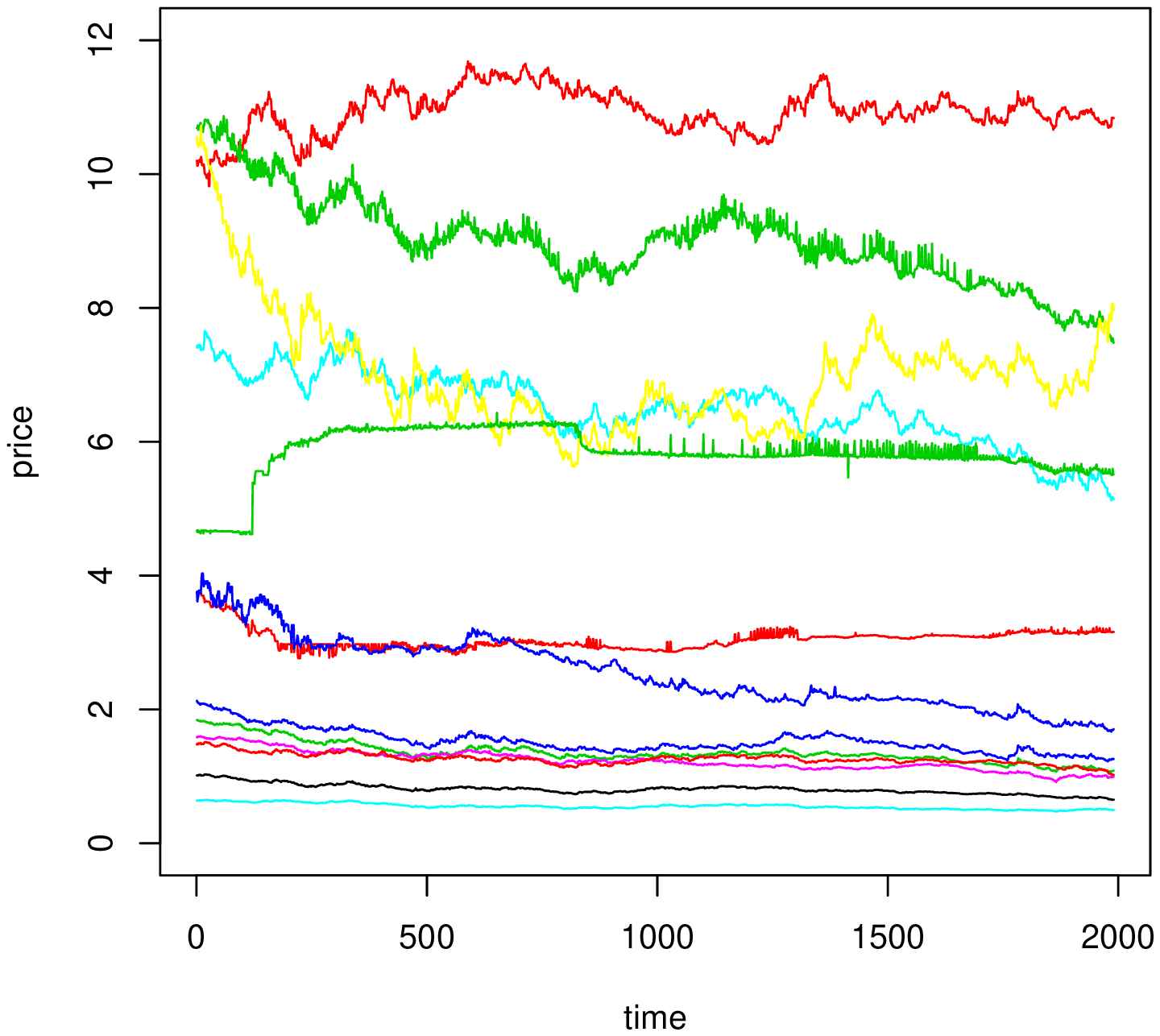}}
\subfigure[DJ Stoxx]
{\includegraphics[width=0.48\textwidth,height=4.2cm]{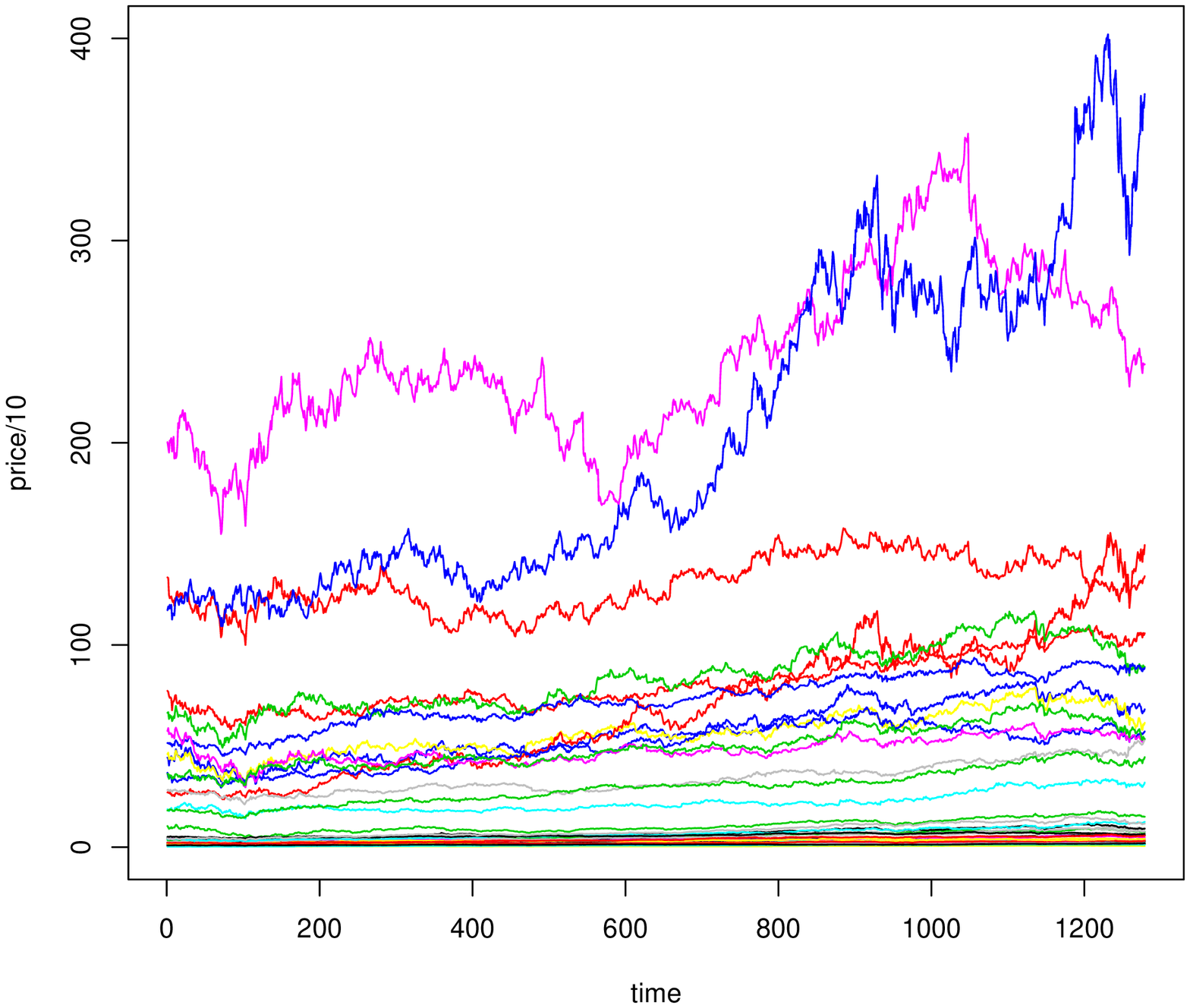}}
\subfigure[Portfolios]
{\includegraphics[width=0.48\textwidth,height=4.2cm]{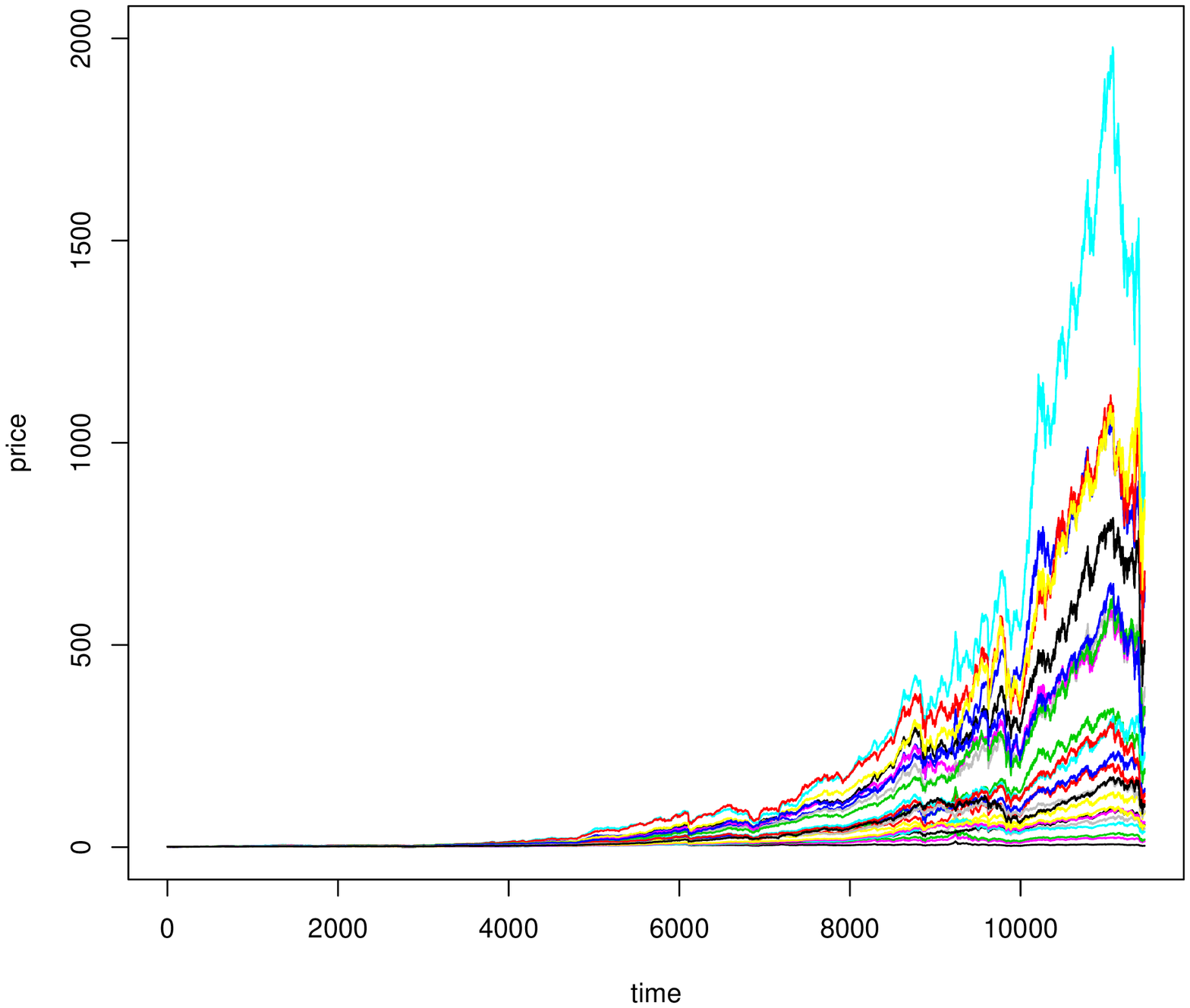}}
\subfigure[FTSE]
{\includegraphics[width=0.48\textwidth,height=4.2cm]{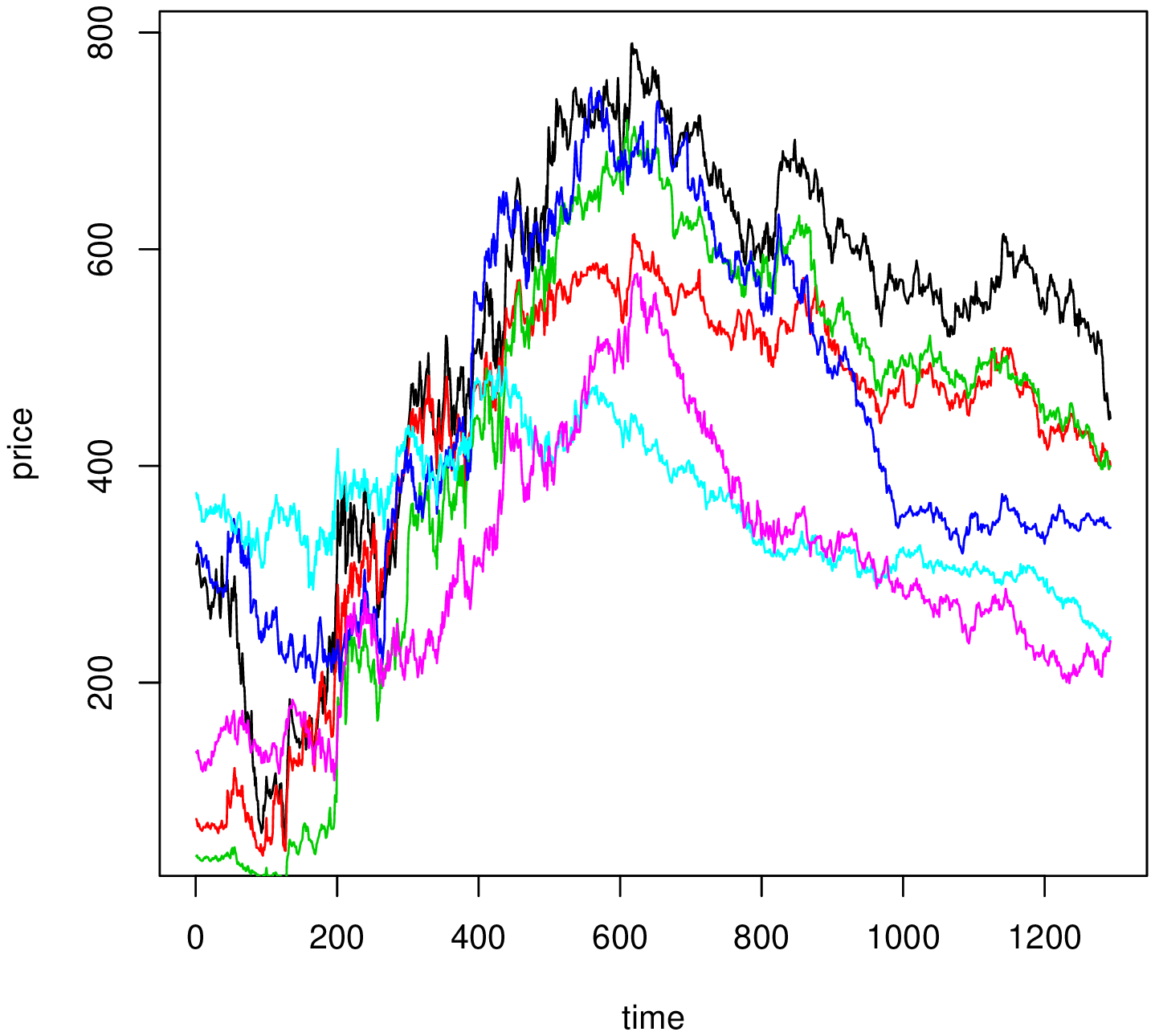}}
\caption{Price Data. Note that the DJ Stoxx prices have been scaled by 10.} \label{fig:FXspotPrices}
\end{figure}

We perform our analysis on 4 datasets; spot Foreign Exchange (FX), constituents of DJ Euro Stoxx, portfolios of NYSE, NASDAQ, AMEX and constituents of FTSE-100 (see Figure \ref{fig:FXspotPrices}).

Our first dataset consists of $19$ spot currencies quoted against the American dollar.
For ease of interpretation, we use the convention ``USD/\ldots", where USD is always the base rate and is read  ``units of foreign currency per 1 USD".
The dataset covers a period of approximately $5\frac{1}{2}$ years of daily data, from $01/10/2002$ until $12/03/2008$. The spot data have been obtained from the ``FXHistory" functionality of OANDA (\verb|www.oanda.com|). 

The second consists of 43 constituents of DJ Euro Stoxx 50, of approximately
$5$ years of closing prices, from  $21/10/2002$ until $13/09/2007$. The data
have been obtained from Yahoo (\verb|http://uk.finance.yahoo.com/|) and have been adjusted
for discontinuities related to financial events, such as stock splits and
bonus issues.

The third dataset are the daily returns on 25 portfolios formed on size and book-to-market from NYSE, NASDAQ and AMEX. The data are from 01/07/63-31/12/08. The data were obtained from 
\verb|http://mba.tuck.dartmouth.edu/pages/faculty/ken.french/|
\verb|data_library.html|.

Our final dataset are 6 constituents (BA, Barclays, Lloyds TSB, M \& S, RBS, Tescos) of the FTSE-100 share index. The daily
data are the adjusted closing prices taken from 17/07/04-17/07/09
and also obtained from Yahoo. 
These particular data will be of interest,
to observe the performance of relatively
simple allocation schemes, during 2 financial crises: the selloff in 2006 caused by algorithmic trading  and the sub-prime mortgage crisis in 2008.
It should be noted that some of our data are clearly subject to survivorship
bias; one should take this into account when looking at the performance measures.

\subsection{The Allocation Strategies}\label{sec:allocStrats}

In our comparison, in addition to the methods developed in Section \ref{sec:R-EWRLS},
we consider 3 standard batch strategies:
\begin{itemize}
\item{NAIVE. This encompasses allocating funds in equal amount to each asset.
As noted in DeMiguel et al.~(2009a), this strategy provides an important benchmark
despite its simplicity.}
\item{Mean-Variance (M-VAR). This is the standard Mean-Variance methodology.
To remove any numerical difficulties with inversion, as noted in introduction, the covariance matrix
is replaced by $\hat{\Sigma} + \vartheta \mathbb{I}_{\gamma\times d}$, where $\vartheta$ is some non-negative constant. The regularisation parameter is chosen as 
$\vartheta=\textrm{trace}(\hat{\Sigma})$}, similar to Ledoit \& Wolf
(2004).
\item{Minimum-Variance (VAR). Standard minimum-variance methodology with the covariance
replaced as for M-VAR.}
\end{itemize}

For R-EWRLS, $\rho(x) = \log\{\cosh(x)\}$. If one could interpret the procedure
as a regression, this would imply a hyperbolic secant error distribution (see Benesty \& Gansler (2001)).
We experimented with more standard choices of $\rho$ (e.g.~Huber's loss function, see Huber (2004)) but
did not find that this significantly affected our conclusions. Note that
we implemented the method of Helmbold et al.~(1998), but did not find a significant
difference with the NAIVE strategy. 

\subsection{Comparison Criteria}\label{sec:compcriteria}

In order to compare and investigate our strategies, we consider various criteria.
The basic idea is to initialize all of the strategies in some way; the first 2 years (504 data points) of each dataset are used for
training (i.e.~omitted afterwards). In particular, and helping to avoid look-ahead
bias, the M-VAR and VAR strategies use the first 2
years of data to estimate the portfolio weights and these are used until the first
re-balancing instant. The re-balancing instant is then determined going forward by the re-balancing window $W$ i.e. re-balancing every 250 data points. Then the data in the time up-to the last re-balancing
period is used to re-estimate the weights.
The weights are initialized
as 1 and are employed from day 2 on-wards. Note
that the actual weights used to compute portfolio returns, they are only based upon
those calculated at re-balancing times. That is to say, we update the weights
for the online methods, but only employ new weights at re-balancing times. As such, trading is infrequent and the transaction cost associated with these allocation strategies is negligible. Therefore, we refrain from using transaction costs, as this would have introduced another layer of assumptions since transaction costs often differentiate substantially from firm to firm given their ``bargaining'' power to negotiate down trading commissions.

The criteria employed are standard in financial applications. The returns for each day are calculated
and we consider: annualized returns and volatility, Sharpe ratio, \% average
daily gain and loss, \% of winning trades (WT), maximum draw-down (MDD) and turnover
(TO).
Of these, perhaps the last 2 need a little explanation. The maximum draw-down
is equal to
$$
-\min\{v_1, v_1+v_2, \dots, v_1+\cdots+v_T, v_2, v_2+v_3, \dots,v_2+\cdots+v_T,\dots,v_T\}
$$
where $v_i$ is the percentage return at period $t$. In words it constitutes the maximum
movement from peak to trough of the cumulative returns, in percentage terms.
The turnover is a measurement of the frequency of trading. It is the average of the absolute difference of the portfolio weights between
re-balancing times. 

\subsection{Initialization}\label{sec:initialization}

\begin{figure}\centering
{\includegraphics[width=\textwidth,height=5cm]{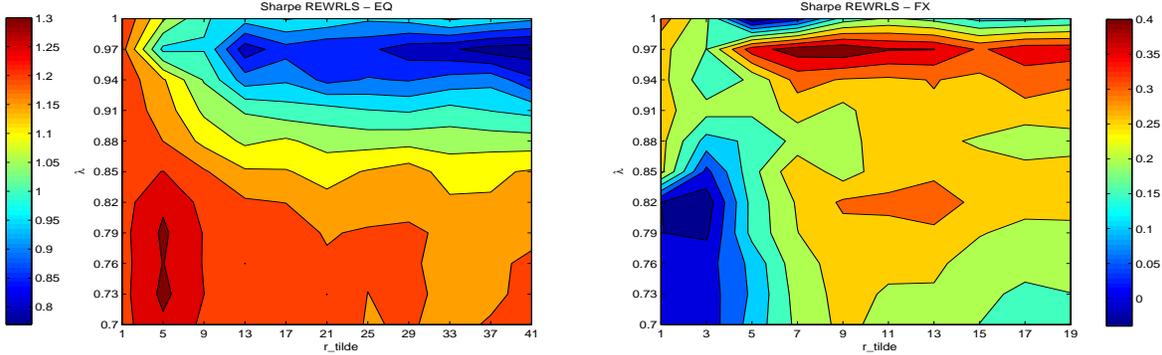}}
\caption{Sharpe ratio to pertubation of the parameters of the R-EWRLS method for DJ Stoxx (left panel) and spot FX
(right panel). The parameter $\lambda$ is the forgetting factor and the parameter $\widetilde{r}$ is the rank of the low rank data approximation. We set $W=250$.} \label{fig:sharpeMatREWRLS}
\end{figure}

We now discuss the selection of parameters for the R-EWRLS
approach.
We explore the Sharpe ratio for the spot FX and DJ Euro Stoxx 50 datasets over a grid of equally-spaced values
of the parameter $\widetilde{r}$ and the forgetting factor $\lambda$.
The results of the exploratory analysis are depicted by means of contour
plots (Figure \ref{fig:sharpeMatREWRLS}).

For the R-EWRLS allocation strategy  using the equities data, we note in Figure \ref{fig:sharpeMatREWRLS} that the Sharpe ratio is positive throughout the parameter space. There is an evident pattern that lower values of $\widetilde{r}$ exhibit higher Sharpe ratio and the difference becomes more pronounced for higher values of $\lambda$.
For the FX dataset, we note (Fig. (\ref{fig:sharpeMatREWRLS})) that there are evident structures of higher Sharpe ratio regions in the parameter space suggesting dependence to the $\lambda$ and $\widetilde{r}$ parameters. In particular, the best performance is achieved for values of $\lambda$ approximately between $0.94$ and $0.98$, when $\widetilde{r}$ is greater than $4$.
On the basis of such plots, we select the values
of $\widetilde{r}$ and $\lambda$.


The choice of $W$ is important for both of our methods and to an extent,
conflicting with $\lambda$. That is to say, instead of making $\lambda$ large,
$W$ can be made smaller and vice-versa. However, the choice of $W$ is also a computational
issue; we may only want to attribute a set memory to the storage associated
to the data. This is the line we follow and set $W=250$ (approximately 1
year of trading) which is not too large for computational purposes and does
not interfere substantially with the data memory profile  implied by $\lambda$, for the purposes of portfolio selection. This is to say that the exponential decay profile would only be truncated for $W$ greater than $250$.
Then the role of $\lambda$ is far clearer with respect to the forgetting of the data.

\subsection{Numerical Results}

The algorithms were run with re-balancing performed
every 50, 150 and 250 days. On the basis of training, the R-EWRLS
used $\widetilde{r}=5$ for the first 2 datasets, $\widetilde{r}=23$
for the third and $\widetilde{r}=2$ for the fourth; respectively
$\lambda\in\{0.8,0.8,0.8,0.75\}$. For O-VAR, $G=100$ (see Section \ref{sec:adaptivedelta})
and the initial $\lambda=0.05$. Note that in each instance, the forgetting
factor converged close to 1 (implying very little forgetting), when there were sufficient re-balancing periods.

We conducted a computational speed comparison between the batch
mean-variance optimization approach against our methods.
We coded the methodologies in  Matlab (version 7.4). 
In a data matrix of $1000\times 500$ dimension, we found that
an iteration needs approximately $15$ milliseconds compared to $2$ seconds for the batch mean-variance computation. In a separate experiment, we increased the number of rows from $1000$ to $5000$. The batch approach computation time increased to $6$ seconds.
The results can be found in Tables \ref{tab:res50}-\ref{tab:res250}.
Some of the annualized volatilities of the strategies exhibited on the tables could be rather high and unrealistic for an investor, but the results are clearly valid as we compare the Sharpe ratio which adjusts for volatility of the underlying strategy. However, one needs to be cautious when comparing maximum draw-down of allocation strategies, as this depends on the volatility of the underlying strategy. 
Let us consider each dataset in turn.

\subsubsection{Spot FX Results}\label{sec:spotFXrates}

Our first observation is that only the R-EWRLS method is consistently 
producing positive returns. Indeed, this is true with respect to different re-balancing
periods. This partly suggests -- inferring from Figure \ref{fig:sharpeMatREWRLS} as well-- that for particularly noisy data, the truncated SVD has a beneficial outcome in the portfolio weights computation; this is in contract to M-VAR whose performance is sensitive to change in $W$ and that result is in line with those reported in the literature. We also note that the portfolio weights of R-EWRLS are more ``active'', as indicated by turnover. This could also imply the R-EWRLS adapts better in the underlying environment, given that delivers consistently better performance than M-VAR.  
For the O-VAR, due to its similarity to the NAIVE
strategy, it is unable to provide positive returns; the latter it exhibits particularly
bad performance here. This is because the NAIVE allocation strategy implies only long positions and is
expected to benefit from a long-term growth typically exhibited in equities,
but not necessarily for FX spot prices.

\subsubsection{DJ Euro Stoxx 50 Results}\label{sec:spotFX}

Moving to the second dataset a more familiar pattern (i.e.~as is often
reported in the literature) is displayed. The NAIVE and VAR strategies perform
relatively well, with quite favourable Sharpe ratios, given the simplicity
of the strategies. The O-VAR method performs marginally better than the VAR
strategy, but with a noticeable increase in turnover. R-EWRLS also delivers satisfactory performance and outperforms M-VAR.

\subsubsection{Portfolio Data}
The portfolio data provide some very interesting results. In this case the
O-VAR provides the most impressive results from a financial perspective,
but performance tends to decrease as the re-balancing time increases. The success of the
O-VAR method is linked to a wide variety of factors. Firstly, due to its
similarity to NAIVE, this method is likely to fair
very well; see Figure \ref{fig:FXspotPrices} 
and the remarks in Section \ref{sec:spotFXrates}. Secondly, O-VAR method should fair well because all parameters are adaptive to the data. However, we note that R-EWRLS is only trained on the first 2 years of data. Since
the data are 45 years long, 2 years is clearly insufficient in which to train
the algorithm. Although this is a little unfair (e.g.~the parameters can be retrained
every 5 years, as would be the case in practice), it highlights a small
deficiency of the R-EWRLS method. Thirdly, against the VAR method, the the smoothness of the portfolio weights is regulated by the Tikhonov regularization.
This may have beneficial outcome in the performance through better estimation (see introduction for rank deficiency discussion). 

\subsubsection{FTSE-100}

The final data provide an interesting set of results. Due to a variety of
economic, cultural (business-wise) and investor related factors, many quantitative
equity hedge funds have performed poorly during the current financial crisis. As
a result, it is of interest from an applied perspective to observe the results
of our models in such a difficult trading period. Rather unsurprisingly,
many of the strategies perform badly. However, in 2 instances,
both of our online methods provide positive returns. This is encouraging,
as to an extent it suggests that the ability to process data as it arrives
and adapt our strategies accordingly is more useful in practice than standard
batch methods.

\subsubsection{General Comments}

On the basis of our investigations, we make the following observations:
\begin{enumerate}
\item{The R-EWRLS method can be successful (positive returns) for noisy data.
However, when the initial training period is insufficient/unreliable,
very unstable results are obtained. In addition, high turnovers were observed
for this method.}
\item{The online and adaptive nature of the O-VAR method, coupled with its
link to the NAIVE strategy leads consistently to strong performance in comparison to the methods tested here.}
\end{enumerate}

In terms of the first point, the R-EWRLS approach is related to M-VAR procedures,
which can work well when there is detectable drift signal in the data. When combined
with the robust scale computation and noise reduction a potentially superior
method is derived. However, there are a number of free parameters, which
are to be set. As a
result, significant training is required and hence the success of the
method is reliant on this latter procedure. 

The second point is clearly reflected in the Tables \ref{tab:res50}-\ref{tab:res250}.
The drawbacks of the R-EWRLS method are alleviated, but with the potential deficiency
of being related to the NAIVE strategy, that is making a naive assumption for the direction of the market by having  long only positions. This can lead to poor performance,
e.g.~for the FX spot data.

\begin{table}\centering
\begin{tabular}{l|l||r|r|r|r|r|r|r|r}
\hline
\hline
Data & Method & \% gain & \% loss & MDD & \% WT & TO &
Ann.R. & Ann.V. & Sharpe\\
\hline
\hline
1& O-VAR&0.17 & -0.17 & 14.50 & 48.08 & 0.59 &-1.34 & 4.08 & -0.33\\\hline
& VAR& 0.20 & -0.22 & 18.09 & 49.70 & 0.074 &-1.94 & 4.58 & -0.43\\\hline
& R-EWRLS & 0.49 & -0.51 & 8.59 & 53.10 & 19.51 &6.22  & 10.90 &0.57\\\hline
& M-VAR & 1.52 & -1.65 & 221.28 & 48.96 & 13.98 &-23.22 & 44.93 &-0.52\\\hline
 & NAIVE & 0.22 & -0.23 & 17.89 & 49.43 & $-$ & -1.89 & 4.91 & -0.38\\\hline
\hline
\hline
2 & O-VAR&0.55 & -0.54 & 9.93 & 55.35 & 0.88 &15.18 & 11.35 &1.34\\\hline
& VAR&0.58 & -0.58 & 10.78 & 55.41 & 0.08 &15.14 & 12.40 &1.26\\\hline
& R-EWRLS & 0.62 & -0.68 & 12.91 & 56.23 & 168.26 &13.82 & 13.63 &1.01\\\hline
& M-VAR & 1.67 & -1.53 & 49.44 & 49.36 & 10.23 & 11.98 & 34.41 &0.35\\\hline
 & NAIVE&0.60&-0.59& 11.10 & 54.63 & $-$ & 16.12 & 12.72 & 1.27\\\hline
\hline
\hline
3 & O-VAR&0.38 & -0.38 & 52.89 & 58.81 & 1.70 & 16.16 & 8.90 &1.82\\\hline
& VAR&0.56 & -0.63 & 67.90 & 56.94 & 0.08 &12.11 & 13.92 &0.87\\\hline
& R-EWRLS & 4.70 & -4.00 & 1281.34 & 50.76 & 73.96 &107.16 & 428.92 &0.25\\\hline
& M-VAR & 1.58 & -1.61 & 240.56 & 55.96 & 12.40 & 43.88 & 42.01 & 1.04\\\hline
 & NAIVE & 0.58 & -0.65 & 67.86 & 56.67 & $-$ & 11.93 & 14.30 &0.84\\\hline
\hline
\hline
4 & O-VAR & 1.73 & -1.73 & 65.74 & 49.43 & 0.99 & -4.99 & 41.19 &-0.12\\\hline
& VAR&1.72 & -1.18 & 139.89 & 50.19 & 0.09 &-11.50 & 43.66 &-0.26\\\hline
& R-EWRLS & 9.27 & -8.53 & 890.60 & 47.91 & 15.15 &-1.06 & 342.45 & -0.003\\\hline
& M-VAR & 2.37 & -2.38 & 182.45 & 47.78 & 4.56 & -28.32 & 55.01 & -0.51\\\hline
 & NAIVE & 1.84 & -1.95 & 155.03 & 50.32 & $-$ & -12.08 & 47.51 & -0.25\\\hline
\hline
\hline
\end{tabular}
\caption{Algorithm performance across datasets. The portfolios are re-balanced every 50 days. See Section \ref{sec:compcriteria} for details.}
\label{tab:res50}
\end{table}

\begin{table}\centering
\begin{tabular}{l|l||r|r|r|r|r|r|r|r}
\hline
\hline
Data & Method & \% gain & \% loss & MDD & \% WT & TO &
Ann.R. & Ann.V. & Sharpe\\
\hline
\hline
1& O-VAR&0.21 & -0.22 & 19.21 & 49.02 & 0.28 &-2.13 & 4.80 & -0.44\\\hline
& VAR& 0.21 & -0.22 & 18.96 & 49.56 & 0.07&-2.11 & 4.66 & -0.45\\\hline
& R-EWRLS & 0.59 & -0.62 & 21.40 & 53.43 & 30.45 &5.69  & 13.62 &0.42\\\hline
& M-VAR & 0.64 & -0.67 & 24.67 & 51.11 & 5.55 &0.89 & 14.87 &0.06\\\hline
\hline
\hline
2 & O-VAR & 0.59 & -0.57 & 10.10 & 54.84 & 0.40 &16.22 & 12.30 &1.31\\\hline
& VAR&0.58 & -0.58 & 10.77 & 55.28 & 0.07 &15.92 & 12.41 &1.28\\\hline
& R-EWRLS & 0.75 & -0.83 & 15.11 & 56.42 & 42.47 &17.03 & 16.74 &1.02\\\hline
& M-VAR & 1.02 & -1.01 & 39.79 & 52.84 & 5.07 & 15.33 & 22.98 &0.67\\\hline
\hline
\hline
3 & O-VAR& 0.40 & -0.43 & 67.96 & 59.23 & 2.40 & 16.54 & 10.00 &1.65\\\hline
& VAR & 0.56 & -0.63 & 67.84 & 56.86 & 0.07 &12.15 & 13.97 &0.87\\\hline
& R-EWRLS & 8.41 & -7.14 & 2238.43 & 50.84 & 133.83 &193.41 & 662.89 &0.29\\\hline
& M-VAR & 1.06 & -1.11 & 253.93 & 56.69 & 7.31 & 31.04 & 31.03 & 0.97\\\hline
\hline
\hline
4 & O-VAR & 2.24 & -2.29 & 97.52 & 51.33 & 1.70 & 9.90 & 62.05 &0.16\\\hline
& VAR&1.75 & -1.83 & 140.23 & 49.81 & 0.07 &-11.25 & 44.08 &-0.26\\\hline
& R-EWRLS & 5.53 & -5.19 & 357.94 & 47.91 & 13.11 &-13.37 & 161.49 & -0.08\\\hline
& M-VAR & 1.82 & -1.76 & 89.99 & 49.05 & 2.21 & -0.01 & 40.32 & -0.02\\\hline
\hline
\hline
\end{tabular}
\caption{Algorithm performance across datasets. The portfolios are re-balanced every 150 days. See Section \ref{sec:compcriteria} for details.}
\label{tab:res150}
\end{table}

\begin{table}\centering
\begin{tabular}{l|l||r|r|r|r|r|r|r|r}
\hline
\hline
Data & Method & \% gain & \% loss & MDD & \% WT & TO &
Ann.R. & Ann.V. & Sharpe\\
\hline
\hline
1& O-VAR&0.21 & -0.22 & 20.07 & 49.49 & 0.33 & -2.30 & 4.66 & -0.49\\\hline
& VAR& 0.21 & -0.22 & 18.68 & 49.63 & 0.07 &-2.04 & 4.71 & -0.43\\\hline
& R-EWRLS & 0.45 & -0.48 & 11.66 & 53.67 & 16.48 & 5.60 &10.00 & 0.56 \\\hline
& M-VAR & 0.52 & -0.53 & 35.83 & 49.43 & 4.97 & -0.35 & 12.11 &-0.29\\\hline
\hline
\hline
2 & O-VAR & 0.57 & -0.57 & 9.97 & 55.22 & 0.71 &16.23 & 12.08 &1.34\\\hline
& VAR&0.59 & -0.57 & 10.70 & 55.03 & 0.06 &15.90 & 12.42 &1.28\\\hline
& R-EWRLS  & 0.74 & -0.79 & 14.46 & 55.85  & 26.79 & 17.60 & 16.08
& 1.10\\\hline
& M-VAR & 0.91 & -0.91 & 22.92 & 53.99 & 3.57 & 19.29 & 19.51 &0.99\\\hline
\hline
\hline
3 & O-VAR& 0.44 & -0.46 & 60.40 & 58.21 & 2.43 & 15.87 & 10.70 &1.48\\\hline
& VAR & 0.57 & -0.64 & 67.57 & 56.91 & 0.08 & 12.11 & 13.99 &0.87\\\hline
& R-EWRLS & 1.98 & -1.93 & 216.25 & 52.04 & 29.89 &25.45 & 58.39 &0.44\\\hline
& M-VAR & 0.86 & -0.93 & 111.61 & 55.99 & 5.91 & 18.37 & 22.14 & 0.83\\\hline
\hline
\hline
4 & O-VAR & 1.71 & -1.88 & 155.03 & 50.95 & 1.21 & -12.00 & 43.27 &-0.28\\\hline
& VAR & 1.71 & -1.87 & 146.86 & 49.81 & 0.09 &-12.11 & 45.47 &-0.27\\\hline
& R-EWRLS & 3.26 & -2.85 & 151.37 & 48.42 & 3.77 &26.84 & 72.74 & 0.37\\\hline
& M-VAR & 1.68 & -1.63 & 77.15 & 48.92 & 1.61 & -3.60 & 37.18 & -0.10\\\hline
\hline
\hline
\end{tabular}
\caption{Algorithm performance across datasets. The portfolios are re-balanced every 250 days. See Section \ref{sec:compcriteria} for details.}
\label{tab:res250}
\end{table}

\section{Summary}\label{sec:conclusion}

We have derived two efficient methods to compute portfolio weights online without the
need of matrix inversion. We compared the two methods with existing techniques
in portfolio optimization using 4 datasets. We
showed that our strategies predominantly outperform the benchmarks,
when performance is measured by Sharpe ratio (note that this includes the
method of Helmbold et al.~(1998)). 


Future research can focus in extending our approach to include transaction
costs (bid-ask spread and commission) as a function of the portfolio weight,
as well as to consider adaptive re-balancing strategies (e.g.~Baltutis, 2009).
For example, the O-VAR method does not explicitly incorporate previous weights
in its estimate and, as such, can lead to high turnovers.
In addition, future work could be focused upon making R-EWRLS fully adaptive. This requires
the online selection of the number of singular values 
and lies on the interface of statistics, finance, signal processing and computer
science. Finally, one of the drawbacks of the O-VAR method was its relation
to NAIVE allocation strategy. This could be removed, for example using $\mathbb{L}_1-$type constraints
leading to an online lasso (Tibshirani (1996)) method (see e.g.~Anagnostopoulos et al.~(2008)). In this context, as the portfolio weights are required to
sum to one (i.e.~standard path-wise co-ordinate optimization (Friedman et
al.~(2007)) does not apply, we are left with an online quadratic programming problem. To
our knowledge, with the exception of Zhang \& Li (2009), there is little
methodology for this problem; we are currently working towards a solution.
Our work also opens up interesting theoretical questions; e.g.~to investigate
the sensitivity of the portfolio weights (as in DeMiguel \& Nogales (2009)) of online algorithms.

\vspace{0.025 in}

\begin{small}

{\ \nocite{*} \centerline{ REFERENCES}
\begin{list}{}{\setlength{\itemindent}{-0.3in}}

\item
{\sc Agarwal}, A., {\sc Hazan}, E., {\sc Kale}, S. \& {\sc
Schapire}, R. E.~(2006). Algorithms for portfolio management based on the
Newton method. \emph{Proc. 23rd Intl. Conf. Mach. Learning}.

\item
{\sc Anagnostopoulos}, C., {\sc Tasoulis}, D., {\sc Adams}, N. M. \& {\sc
Hand}, D. J.~(2008). Online optimization for variable selection in data streams.
\emph{ECAI 2008, 18th Euro. Conf. Art. Intel.}, 132--136.

\item
{\sc Baltutis}, M.~(2009). Non-stationary stock returns and time to revise
the optimal portfolio. Technical Report, University of Vienna.

\item
{\sc Benesty}, J. \& {\sc Gansler} T.~(2001). A robust fast recursive least squares adaptive algorithm.
\emph{Acoustics, Speech, and Signal Processing .Proceedings (ICASSP'01)}, {\bf 6}.

\item
{\sc Britten-Jones}, M.~(1999). The sampling error in estimates of mean-variance
efficient portfolio weights.
\emph{J. Finan.}, {\bf 54}, 655--671.

\item
{\sc Bunch}, J. R. \& {\sc Nielsen}, J. R.~(1978). Updating the singular
value decomposition. \emph{Numer. Math.}, {\bf 31}, 111--129.

\item
{\sc Chapados}, N. \& {\sc Bengio}, Y.~(2007). Noisy $k$ best paths for approximated
dynamic programming with application to portfolio optimization.
\emph{J. Comput.}, {\bf 2}, 12--19.

\item
{\sc Cipra}, T. \& {\sc Romera}, R.~(1991). Robust Kalman filter and its
application to time series analysis.
\emph{Kybernetika}, {\bf 27}, 481--494.

\item
{\sc DeMiguel}, V., \& {\sc Nogales}, F. J.~(2009). Porfolio selection with
robust estimation.
\emph{Op. Res.}, {\bf 57}, 560--577.

\item
{\sc DeMiguel}, V., {\sc Garlappi}, L.,  \& 
{\sc Uppal}, R.~(2009a). Optimal versus naive diversification: How
inefficient is the $1/N$ strategy.
\emph{Rev. Finan. Stud.}, {\bf 22}, 1915--1953.

\item
{\sc DeMiguel}, V., {\sc Garlappi}, L., {\sc Nogales}, F. J. \& 
{\sc Uppal}, R.~(2009b). A generalized approach to portfolio optimization:
Improving performance by constraining portfolio norms.
\emph{Manage. Sci.}, {\bf 55}, 798--812.

\item
{\sc Deng}, G.~(2008). Sequential and adaptive learning algorithms for $M-$estimation.
\emph{EURASIP J. Adv. Sig. Proc.}, {\bf 2008}, Article ID 459586.

\item
{\sc Fabozzi}, F., {\sc Kolm}, P. N., {\sc Pachamanova}, D. A., {\sc Focardi},
S. M.~(2007). \emph{Robust Portfolio Optimization and Management}. Wiley: Hoboken, NJ.

\item
{\sc Fabozzi}, F., {\sc Huang}, D. \& {\sc Zhou}, G.~(2009). Robust portfolios:
contributions from operations research and finance. \emph{Ann. Op. Res.},
To appear.

\item
{\sc Friedman}, J., {\sc Hastie}, T., {\sc Holfling}, H.  \& {\sc Tibshirani}, R.~(2007). Pathwise coordinate optimization. \emph{Ann. Appl. Statist.}, {\bf 2}, 302--332.

\item
{\sc Frauendorfer}, K., \& {\sc Siede}, H.~(2000). Portfolio selection using
multi-stage stochastic programming. \emph{Centr. Eur. J. Op. Res.}, {\bf 7}, 277--290.

\item
{\sc Hamilton}, J. D.~(1994). \emph{Time Series Analysis}. Princeton University
Press: Princeton.

\item
{\sc Hansen}, P. C.~(1987). The truncated SVD as a method for regularization.
\emph{BIT Numer. Math.}, {\bf 27}, 534--553.

\item
{\sc Hansen}, P. C.~(1996). \emph{Rank-Deficient and Discrete Ill-Posed Problems}.
Polyteknisk Forlag.

\item
{\sc Haykin}, S.~(1996). \emph{Adaptive Filter Theory}. Prentice Hall: New
Jersey.

\item 
{\sc Helmbold}, D. P., {\sc Schapire}, R. E., {\sc Singer}, Y. \& {\sc Warmuth}, M. K.~(1998). 
On-line portfolio selection using multiplicative updates.
\emph{Math. Finance}, {\bf 8}, 325--347.

\item
{\sc Huber}, P. J.~(2004). \emph{Robust Statistics}. Wiley: New York.

\item 
{\sc Jagannathan}, R., \& {\sc Ma}, T.~(2003). Risk reduction in large portfolios:
why imposing the wrong constraints helps. \emph{J. Finance}, {\bf 58}, 1651--1684.

\item 
{\sc Kuhn}, D., {\sc Parpas}, P., {\sc Rustem}, B.  \& {\sc Fonseca}, R.~(2009). Dynamic mean-variance portfolio analysis under model risk. \emph{J. Comp. Finance}, {\bf 12}, 91--115.

\item
{\sc Keehner}, J.~(2007). Milliseconds are focus in algorithmic trades. Reuters. Available from \verb|http://www.reuters.com/|.

\item 
{\sc Ledoit}, O., \& {\sc Wolf}, M.~(2003). Improved estimation
of the covariance matrix of stock returns with an application to portfolio
selection. \emph{J. Empr. Finance}, {\bf 10}, 603--621.

\item 
{\sc Ledoit}, O., \& {\sc Wolf}, M.~(2004). A well-conditioned estimator
for large dimensional covariance matrices. \emph{J. Multivariate Anal.}, {\bf 88}, 365--411.

\item
{\sc Li}, D., \& {\sc Ng}, W. L.~(2000). Optimal dynamic portfolio selection:
Multi-period mean-variance formulation. \emph{Math. Finance}, {\bf 10}, 387-406.

\item
{\sc Markowitz}, H. (1952). Mean-variance analysis in portfolio choice and
capital markets. \emph{J. Finance}, {\bf 7}, 77-91.

\item
{\sc Martin}, R. D.~(1979). Approximate conditional-mean type smoothers and
interpolators. \emph{Proc. IEEE Conf. Dec. Contr}.

\item
{\sc Masreliez}, C.~(1975). Approximate non-Gaussian filtering with linear
state and observation relations. \emph{IEEE Trans. Aut. Contr.}, {\bf 20},
107--110.

\item
{\sc Merton}, R. C.~(1980). On estimating the expected return on a market:
An exploratory investigation. \emph{J. Financial Econ.}, {\bf 8},
323--361.

\item
{\sc Montana}, G., {\sc Triantafyllopoulos}, K. \& {\sc Tsagaris}, T. (2008)
Data stream mining for algorithmic trading.
\emph{Proceedings of the 2008 ACM symposium on Applied computing}, 966--970.

\item
{\sc Montana}, G., {\sc Triantafyllopoulos}, K. \& {\sc Tsagaris}, T. (2009)
Flexible least squares for temporal data mining and statistical
  arbitrage.
\emph{Expert Systems and Applications}. {\bf 36}, 2819--2830.

\item
{\sc Sayed}, A. H.~(2003). \emph{Fundamentals of Adaptive Filtering}. IEEE: New Jersey.

\item
{\sc Schick}, I. C. \& {\sc Mitter}, S. K.~(1994). Robust recursive estimation
in the presence of heavy tailed observation noise. \emph{Ann. Statist.}, {\bf 22}, 1045--1080.

\item
{\sc Smith}, K. F.~(1967). A transition model for portfolio revision. \emph{J. Finance}, {\bf 22}, 425--439.

\item
{\sc Stewart}, G. W.~(1993) On the early history of the singular value decomposition. \emph{Siam Review}, {\bf 35}, 551--566.

\item
{\sc Tibshirani}, R.~(1996). Regression shrinkage and selection via the lasso. \emph{J. R. Statist. Soc. Ser. B}, {\bf 58}, 267--288.

\item
{\sc Tibshirani}, R.~(2008). Fast computation of the median by successive binning. Stanford University. Preprint.

\item 
{\sc Yu}, K. B.~(1991). Recursive updating the eigenvalue decomposition of
a covariance matrix. \emph{IEEE Trans. Sig. Proc.}, {\bf 39}, 1136--1145

\item 
{\sc Zhang}, Y. \& {\sc Li}, Z.~(2009). Zhang neural network for online solution
of time-varying convex quadratic program subject to time varying linear-equality
constraints. \emph{Phys. Lett A}, {\bf 373}, 1639--1643.

\end{list}
}

\end{small}

\end{document}